\documentclass[pra,10pt,aps,twocolumn,preprintnumbers,amsmath,amssymb,longbibliography]{revtex4-1}
\usepackage{graphicx}
\usepackage{dcolumn}
\usepackage{bm}
\usepackage{url}
\begin{document}

\title{Simple scaling rules governing work functions of two-dimensional materials}

\author{James C. Ellenbogen}
\email{ellenbgn@mitre.org}
\author{Sahasra Yellepeddi}
\author{Zev Goldhaber-Gordon}

\affiliation{%
Emerging Technologies Division, The MITRE Corporation, McLean, VA, USA 22102
}

\date{\today}

\begin{abstract}
This paper demonstrates that values of work functions $W$ for a variety of planar and buckled two-dimensional (2D) materials scale linearly as a function of the quantity $1/r_{WS}$, where $r_{WS}$ is the Wigner-Seitz radius for a 2D material.  Simple procedures are prescribed for estimating $r_{WS}$.  Using them, this linear scaling relation, which is founded in electrostatics, provides a quick and easy method for calculating values of $W$ from basic, readily available structural information about the materials.  These easily determined values of $W$ are seen to be very accurate when compared to values from the literature.  Those derive from experiment or from challenging, computationally intensive density-functional-theory calculations.  Values from those sources also conform to the linear scaling rules.  Since values of W predicted by the rules so closely match known values of W, the simple scaling methods described here also are applied to predict $W$ for 2D materials (TiN, BSb, SiGe, and GaP) for which no values have appeared previously in the literature.
\end{abstract}

\pacs{TBD}

\maketitle

\section{\label{sec:Intro}Introduction}

There has been intense research, worldwide, on two-dimensional (2D) materials since the discovery of graphene, with many new types having been synthesized~\cite{miro2014atlas,gjerding2021recent}.  These have the potential to influence every branch of condensed-matter physics and materials science~\cite{,zhao2024prx}, while a number of high-value applications are envisioned and under development~\cite{glavin2020emerging}.  In most cases, applications are dependent upon knowledge of the materials' properties~\cite{gjerding2021recent,shanmugam2022review}---especially electronic properties, such as their electron detachment energies, or work functions $W$.  It has been challenging, though, to determine $W$ experimentally and also  computationally~\cite{momeni2020multiscale}.

\begin{figure}[b]
\begin{center}
\includegraphics[width=0.48\textwidth]{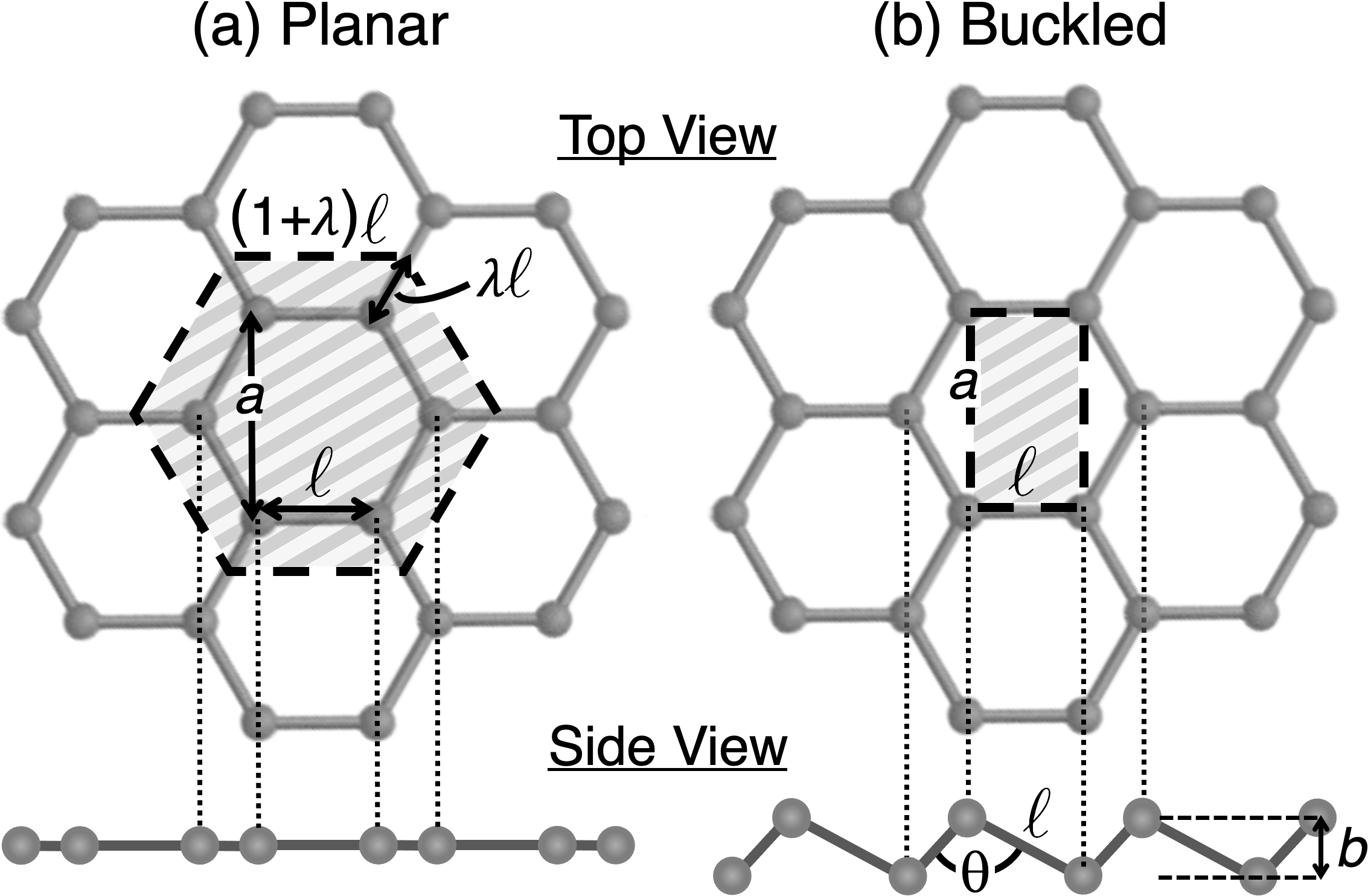}
\end{center}
\caption{\small \small Schematic diagrams of planar and buckled 2D materials, viewed both from the top and from the sides of their surfaces, and showing how the lattice constants $a$, bond lengths $\ell$, bond angle $\theta$, and buckling height $b$ characterize the structures of the two types of materials.  Gray-hashed regions enclosed within the dashed lines are the average areas $A$ in the plane corresponding to above-the-plane volumes $V$ that are estimated here to be available to the mobile $\pi$-electrons from each hexagonal group of atoms.  These electrons form a planar conductive surface producing an image potential that acts upon a departing electron.  Parameter $\lambda$ specifies the fraction of a bond length along which electrons associated with a hexagonal group of atoms delocalize, on average, into the area of a neighboring hexagon.}
\label{fig:2DMaterialsSchematic}
\end{figure}

However, this paper describes the discovery and demonstration of a simple, fundamental regularity that governs values of $W$ for 2D materials having either planar or buckled structures, both of which are depicted schematically in Fig.~\ref{fig:2DMaterialsSchematic}. It is shown that $W$ varies linearly along a scaling line given by the equation
\begin{equation}
\label{eq:basic_W_eqn}
W = \frac{1}{4} \Big(\frac{1}{r_{WS}}\Big) \,,
\end{equation}
where $r_{WS}$ is the Wigner-Seitz radius~\cite{ashcroft1976solid} for a 2D material, and we take its reciprocal to be the scaling variable.

%
%

\begin{figure}[t]
\begin{center}
\includegraphics[width=0.48\textwidth]{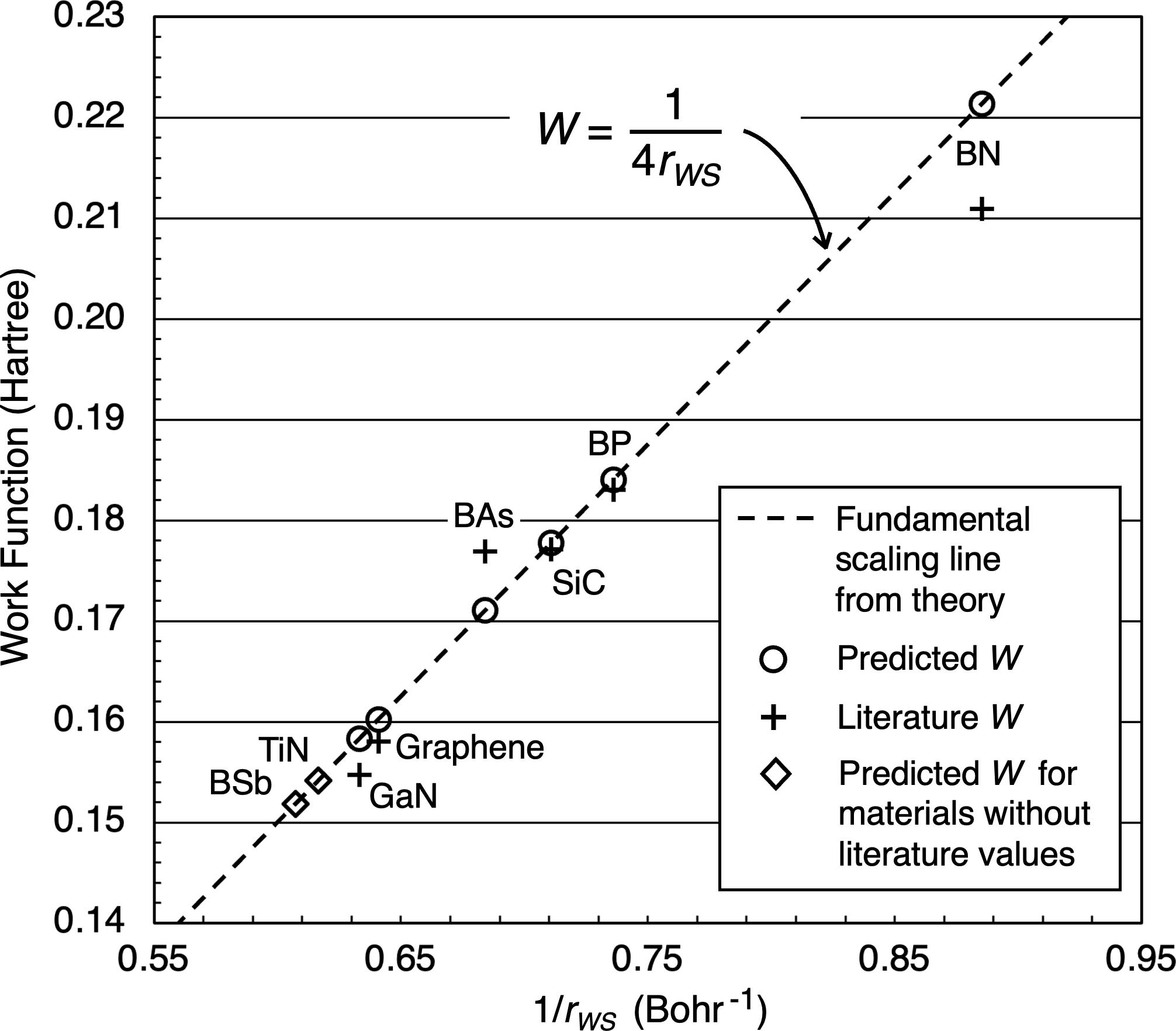}
\end{center}
\caption{\small \small Plot of $W$ versus $1/r_{WS}$ for $planar$ 2D materials.  Values from work function calculations performed here for those materials ($\circ$'s), using the procedures of Section~\ref{sec:PlanarFormulas}, are compared with values from prior calculations and experiments found in the literature ($+$'s).  Points designated by diamonds ($\diamond$'s) represent predictions of $W$ for which no values appear in prior literature.  See part (a) of Table~\ref{tab:Results} for values plotted in this figure.}
\label{fig:PlanarResults}
\end{figure}

\begin{figure}[t]
\begin{center}
\includegraphics[width=0.48\textwidth]{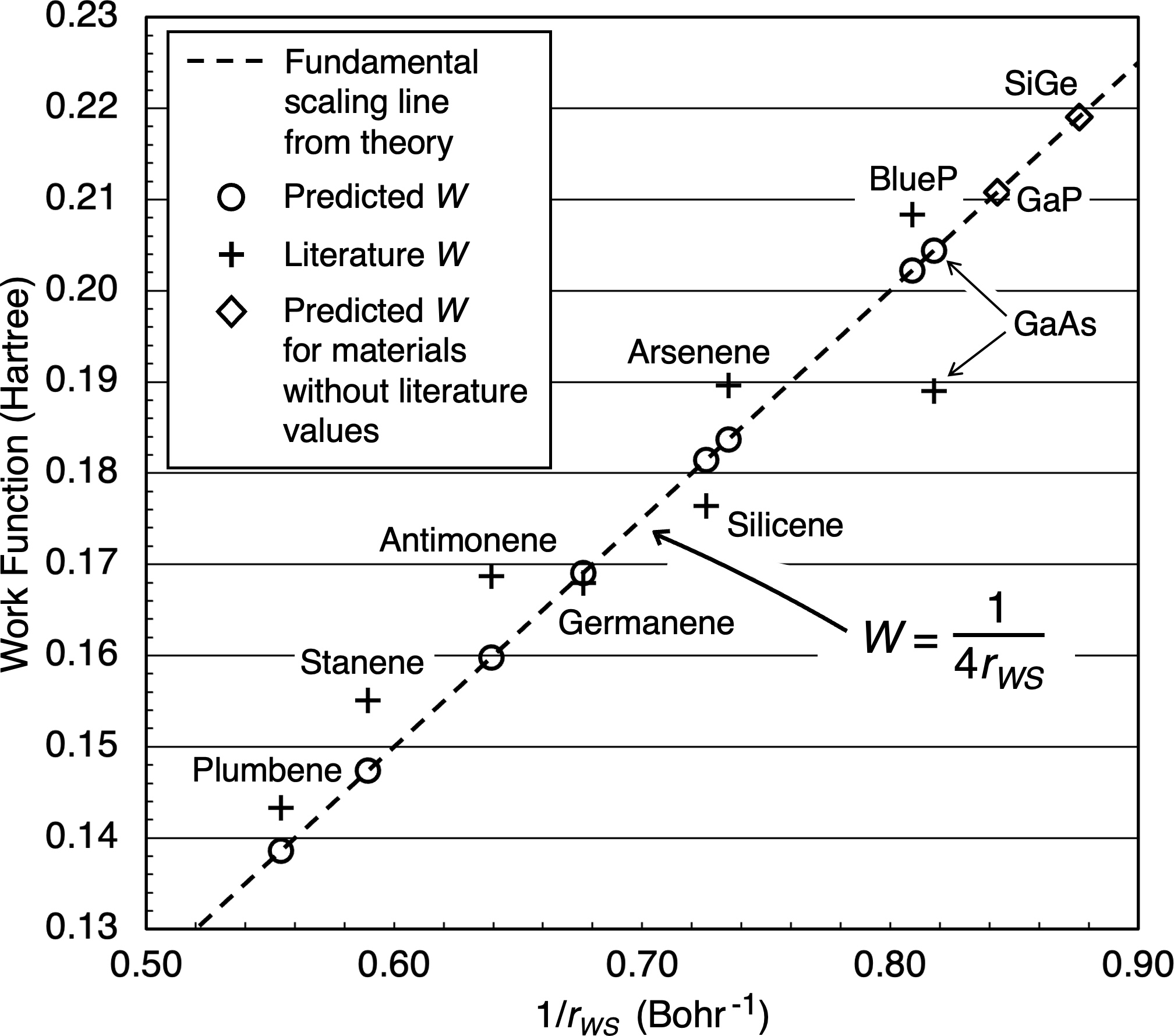}
\end{center}
\caption{\small \small Plot of $W$ versus $1/r_{WS}$ for $buckled$ 2D materials. Values from work function calculations performed here for those materials ($\circ$'s), using the procedures of Section~\ref{sec:FirstApproxn}, are compared with values from prior calculations in the literature ($+$'s). Points designated by diamonds ($\diamond$'s) represent predictions of $W$ for which no values appear in prior literature. See part (b) of Table~\ref{tab:Results} for values plotted in this figure.} 
\label{fig:BuckledResults}
\end{figure}

This fundamental regularity applies to $W$ values that were calculated individually, one-by-one, by prior investigators for a number of such materials, but all fall close to and along scaling lines when plotted versus $1/r_{WS}$.  This is shown explicitly for planar and buckled 2D materials, respectively, in the graphs within Figs.~\ref{fig:PlanarResults} and \ref{fig:BuckledResults}.  Therein, points corresponding to values of $W$ drawn from prior literature are represented by $+$'s, while the dashed line represents $W$ scaling with $1/4r_{WS}$, according to Eq.~(\ref{eq:basic_W_eqn}).

Also, in Tables~\ref{tab:Results} and \ref{tab:Differences} it is seen how small are the differences between literature values of $W$ and those we calculate from the scaling rule of Eq.~(\ref{eq:basic_W_eqn}), which are shown as $\circ$'s right on the dashed scaling lines in the graphs.
To perform these calculations and construct these graphs and tables, we provide simple, physically-grounded methods by which $r_{WS}$ can be estimated with accuracy for 2D materials.  Those methods are detailed below in Section~\ref{sec:Methods}.

The insights summarized above and other elements of this work are products of an electrostatics-based approach that has been applied previously to analyze the work functions of metal surfaces, small metal particles, fullerenes, and other nanostructures~\cite{Wood1981,VanStaveren1987,Perdew1988,atanasov_and_ellenbogenPRA2017}.  That approach is adapted here to treat 2D materials.

Equation~(\ref{eq:basic_W_eqn}) arises from the electrostatic treatment of a 2D material as a conducting plane of effectively infinite extent.  As is well known, an image-charge model of such a conducting plane~\cite{jackson_electrdynamics2021,Landau_and_Lifschitz1984} prescribes that the energy in atomic units required to remove an electron from a distance $\delta$ above the plane to an infinite distance away is $1/4\delta$.  The enabling concept introduced here to obtain Eq.~(\ref{eq:basic_W_eqn}) is that we may take
\vspace{-1mm}
\begin{equation}
\label{eq:delta_eqn}
\delta = r_{WS} \, .
\end{equation}
\vspace{-5mm}

We arrive at this valuation for $\delta$ by recognizing that, in a gas of mobile electrons, the Wigner-Seitz radius~\cite{ashcroft1976solid} is the average radius of a sphere containing just one electron.  Then, assuming the valence electrons above a 2-D material are approximately uniformly distributed in a single layer across the material's surface, $r_{WS}$ is an approximate measure of the average extent of the bound electron distribution in the direction perpendicular to the surface---i.e., the direction in which a departing electron travels.  Thus, it is logical to estimate the average distance $\delta$ above the surface from which a detached electron departs to be $r_{WS}$~\cite{delta_note}.

%
%

%
%

%
\begin{table*}[t]
\caption{\small \small Results from 2D material work function calculations performed in this work, using the equations listed in the table, are compared with values from the literature.  Values for $r_{WS}$ given in column 2 of the table have been obtained using dimensional parameters presented within Table~\ref{tab:Parameters} and the methods described in Section~\ref{sec:Methods} of this work.  Literature sources of $W$ are cited in the last column of the table.  Except for the literature value of $W$ for graphene, all the literature values of $W$ in column 8 of the table derive from DFT calculations. Materials designated by an asterisk (*) have not yet had a value for $W$ appear in prior literature.  Thus the values listed here represent the first predictions for their work functions. }
\label{tab:Results} 
\vspace{1mm} 
\begin{ruledtabular}
\begin{tabular}{cccccccccc}
  &  &  &  &Effective       &  &    &     &  &  \\
  &  &  &  &Radius of          &Out-of-Plane  &    &     &  &  \\
  &  &  &  &Out-of-Plane  &Curvature  &${W}$  &${W}$  &  &  \\
  &  &  &  &Curvature\footnotemark[1],  &Correction,  &Predicted in  &From  &Pct.  &Refs.  \\
2D Material  &$r_{WS}$  &1/($r_{WS}$)  &1/(4$r_{WS}$)  &$R_{eff}$  &(3/8)(1/$R_{eff}$)  &This Work  &Literature  &Difference  &for ${W}$  \\
  &(Bohr)  &(Bohr$^{-1}$)  &(Hartree)  &(Bohr)  &(Hartree)  &(Hartree)  &(Hartree)  &  &  \\
\hline
\vspace{-2.5mm} \\
\multicolumn{10}{c}{\underline{(a) For Planar Structures Using Eq. (\ref{eq:basic_W_eqn})}}  \\
\vspace{-3mm} \\
Graphene  &1.561  &0.641  &0.160  &\textendash  &\textendash  &0.160  &0.158  &1.3  &\cite{rut2020graphene}  \\
2D BAs  &1.461  &0.684  &0.171  &\textendash  &\textendash  &0.171  &0.1769  &-3.4  &\cite{shahriar2022vacancy} \\
2D BN  &1.129  &0.885  &0.221  &\textendash  &\textendash  &0.221  &0.211  &4.7  &\cite{thomas2020strain}  \\
2D SiC  &1.407  &0.711  &0.178  &\textendash  &\textendash  &0.178  &0.177  &0.6  &\cite{zhang2022electronic}  \\
2D BP  &1.358  &0.736  &0.184  &\textendash  &\textendash  &0.184  &0.183  &0.5  &\cite{hasan2020strain}  \\
2D GaN  &1.579  &0.633  &0.158  &\textendash  &\textendash  &0.158  &0.155  &1.9  &\cite{cui2018alkali}  \\
2D TiN*  &1.622  &0.617  &0.154  &\textendash  &\textendash  &0.154  &\textendash  &\textendash  &\textendash  \\
2D BSb*  &1.647  &0.607  &0.152  &\textendash  &\textendash  &0.152  &\textendash  &\textendash  &\textendash  \\
\vspace{-2.5mm} \\
\multicolumn{10}{c}{\underline{(b) For Planar Model of Buckled Structures Using Approximation 1, Eq. (\ref{eq:basic_W_eqn})}}  \\
\vspace{-3mm} \\
Silicene  &1.378  &0.726  &0.181  &\textendash  &\textendash  &0.181  &0.18  &2.8  &\cite{qin2012first}  \\
Germanene  &1.479  &0.676  &0.169  &\textendash  &\textendash  &0.169  &0.168  &0.6  &\cite{chen2016electronic}  \\
Stanene  &1.696  &0.589  &0.147  &\textendash  &\textendash  &0.147  &0.155  &-5.2  &\cite{chen2016electronic2,garg2017band}  \\
Plumbene  &1.804  &0.554  &0.139  &\textendash  &\textendash  &0.139  &0.143  &-2.8  &\cite{jamwal2023tailoring}  \\
Antimonene  &1.565  &0.639  &0.160  &\textendash  &\textendash  &0.160  &0.169  &-5.3  &\cite{kripalani2018strain}  \\
Arsenene  &1.361  &0.735  &0.184  &\textendash  &\textendash  &0.184  &0.190  &-3.2  &\cite{sun2018exceptional}  \\
BlueP\footnotemark[2]  &1.236  &0.809  &0.202  &\textendash  &\textendash  &0.202  &0.208  &-2.9  &\cite{zhao2015new}  \\
2D GaAs  &1.223  &0.818  &0.204  &\textendash  &\textendash  &0.204  &0.189  &7.9  &\cite{shahriar2022adsorption}  \\
2D SiGe*  &1.141  &0.876  &0.219  &\textendash  &\textendash  &0.219  &\textendash  &\textendash  &\textendash  \\
2D GaP*  &1.186  &0.843  &0.211  &\textendash  &\textendash  &0.211  &\textendash  &\textendash  &\textendash  \\
\vspace{-2.5mm} \\
\multicolumn{10}{c}{\underline{(c) For Nonplanar Model of Buckled Structures Using Approximation 2, Eq. (\ref{eq:Corrected_W_Eqn})}}  \\
\vspace{-3mm} \\
Silicene  &2.081  &0.481  &0.120  &6.899  &0.054  &0.175  &0.18  &-0.6  &\cite{qin2012first}  \\
Germanene  &2.243  &0.446  &0.111  &6.867  &0.055  &0.166  &0.168  &-1.2  &\cite{chen2016electronic}  \\
Stanene  &2.584  &0.387  &0.097  &7.851  &0.048  &0.145  &0.155  &-6.5  &\cite{chen2016electronic2,garg2017band}  \\
Plumbene  &2.754  &0.363  &0.091  &7.852  &0.048  &0.139  &0.143  &-2.8  &\cite{jamwal2023tailoring}  \\
Antimonene  &2.514  &0.398  &0.099  &5.405  &0.069  &0.169  &0.169  &0.0  &\cite{kripalani2018strain}  \\
Arsenene  &2.172  &0.460  &0.115  &4.905  &0.076  &0.192  &0.190  &1.1  &\cite{sun2018exceptional}  \\
BlueP\footnotemark[2]  &1.970  &0.508  &0.127  &4.513  &0.083  &0.210  &0.208  &1.0  &\cite{zhao2015new}  \\
2D GaAs  &1.854  &0.539  &0.135  &6.90  &0.054  &0.189  &0.189  &0.0  &\cite{shahriar2022adsorption}  \\
2D SiGe*  &1.727  &0.579  &0.144  &6.84  &0.055  &0.200  &\textendash  &\textendash  &\textendash  \\
2D GaP*  &1.792  &0.558  &0.140  &6.81  &0.055  &0.195  &\textendash  &\textendash  &\textendash  \\  \\
\vspace{-11mm} \\
\footnotetext[1]{Values for $R_{eff}$ have been obtained in the calculations presented in Table~\ref{tab:EffectiveRadius}, using the set of parameter values presented there and the method described in Section~\ref{sec:SecondApproxn}.}
\footnotetext[2]{BlueP stands for blue phosphorene, which is the buckled form of 2D phosphorus.}
\end{tabular}
\end{ruledtabular}
\end{table*}
%

%
\begin{table}[t]
\caption{\small \small Assessment of differences and root-mean-square (RMS) differences between values of $W$ from the theory of this work and those from the literature.  Means and differences given in this table are determined from values of W presented in Table~\ref{tab:Results}.  Observe that all the energy differences given below in eV, except the absolute difference for GaAs in part (b), are less than 0.3 eV.  That has been estimated~\cite{DeWaele_etal2016} to be the approximate error for DFT calculations of $W$, which are the source for most of the literature values used here.  Thus, the values of $W$ calculated much more simply in this work and presented in Table~\ref{tab:Results} are within that error and might be regarded as being as accurate as those from DFT.}
\label{tab:Differences}
\begin{ruledtabular}
\begin{tabular}{cccccc}
& Mean of & \multicolumn{2}{c}{Absolute} & \multicolumn{2}{c}{RMS}  \\
 & All & \multicolumn{2}{c}{Differences} &  \multicolumn{2}{c}{Difference} \\
 & Literature & \multicolumn{2}{c}{Between Lit.} & \multicolumn{2}{c}{Btwn.\!\! All  Lit.}\\
2D & Values  & \multicolumn{2}{c}{\& Theory} &  \multicolumn{2}{c}{\& Theory} \\
Material & for W &  \multicolumn{2}{c}{\underline{Values for W}} & \multicolumn{2}{c}{\underline{Values for W}} \\
\vspace{-2.5mm} \\
 & (Hartree) & (Hartree) & (eV) & (Hartree) & (eV) \\
\hline
\vspace{-2.5mm} \\
\multicolumn{6}{c}{\underline{(a) For Planar Structures Via Eq. (\ref{eq:basic_W_eqn})}} \\
\vspace{-3mm} \\
Graphene  &    &  0.002  &  0.054  &    &   \\
2D BAs  &    &  0.006  &  0.163  &    &   \\
2D BN  &  0.177  &  0.010  &  0.272  &  0.005  &  0.14 \\
2D SiC  &    &  0.001  &  0.027  &    &   \\
2D BP  &    &  0.001  &  0.027  &    &   \\
2D GaN  &    &  0.003  &  0.082  &    &   \\
\vspace{-2mm} \\
\multicolumn{6}{c}{\underline{(b) For Planar Model of Buckled Structures, Eq. (\ref{eq:basic_W_eqn})}}  \\
\vspace{-3mm} \\
Silicene  &    &  0.005  &  0.136  &    &   \\
Germanene  &    &  0.001  &  0.027  &    &   \\
Stanene  &    &  0.008  &  0.218  &    &   \\
Plumbene  &  0.175  &  0.004  &  0.109  & 0.005 & 0.14 \\
Antimonene  &    &  0.009  &  0.245  &    &   \\
Arsenene  &    &  0.006  &  0.163  &    &   \\
BlueP  &    &  0.006  &  0.163  &    &   \\
2D GaAs  &    &  0.015  &  0.408  &    &   \\
\vspace{-2mm} \\
\multicolumn{6}{c}{\underline{(c) For Nonplanar Model of Buckled Structures, Eq. (\ref{eq:Corrected_W_Eqn})}}  \\
\vspace{-3mm} \\
Silicene  &    &  0.001  &  0.027  &    &   \\
Germanene  &    &  0.002  &  0.054  &    &   \\
Stanene  &    &  0.010  &  0.272  &    &   \\
Plumbene  &  0.175  &  0.004  &  0.109  & 0.004  & 0.11  \\
Antimonene  &    &  0.000  &  0.000  &    &   \\
Arsenene  &    &  0.002  &  0.054  &    &   \\
BlueP  &    &  0.002  &  0.054  &    &   \\
2D GaAs  &    &  0.000  &  0.000  &    &   \\
\vspace{-4mm} \\
\end{tabular}
\end{ruledtabular}
\end{table}
%


%
\begin{figure}
\begin{center}
\includegraphics[width=0.48\textwidth]{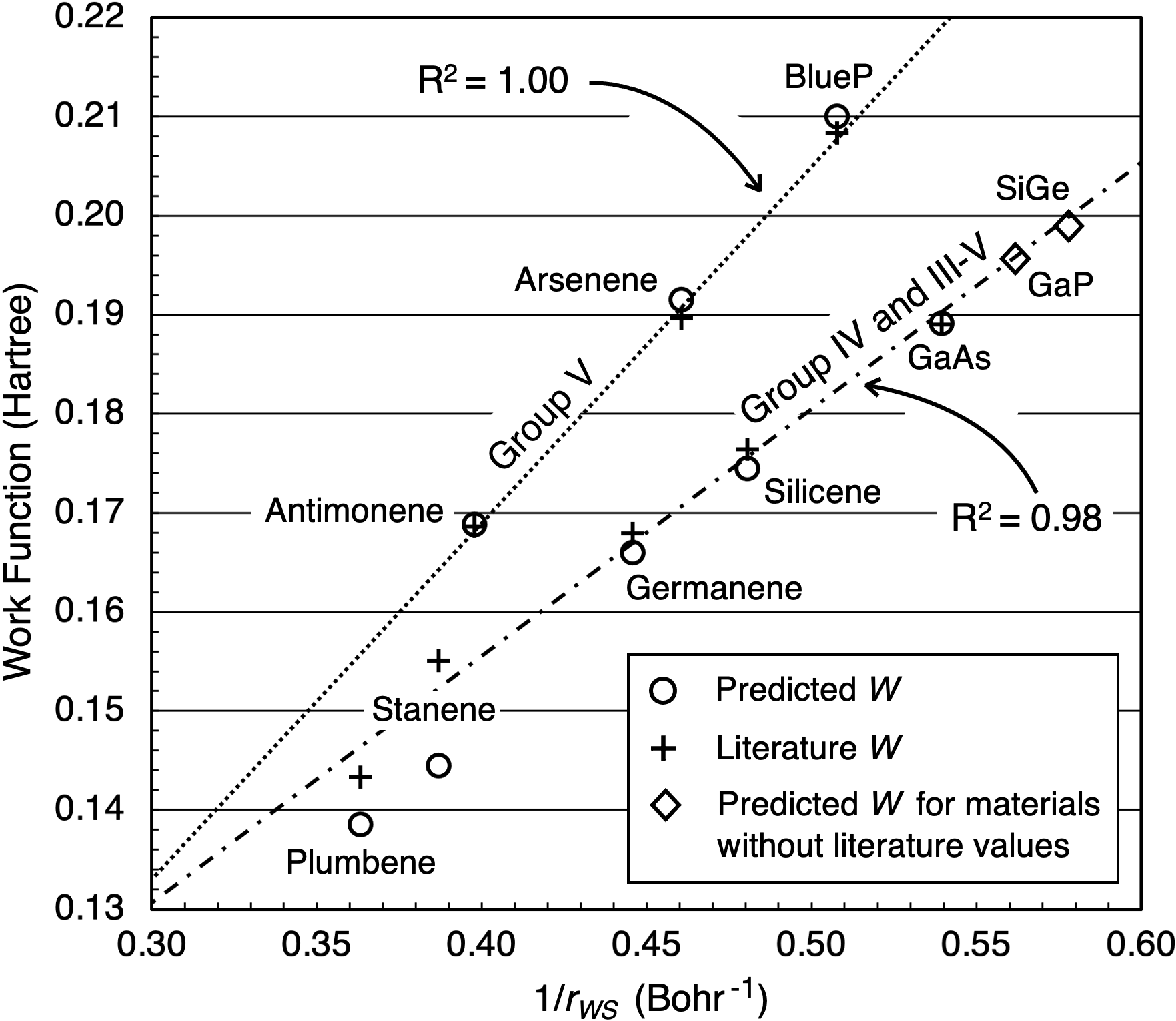}
\end{center}
\caption{\small \small Results from work function calculations performed here ($\circ$'s) using the the $non$planar approximation of Section~\ref{sec:SecondApproxn} for $buckled$ 2D materials are compared with values from prior calculations described in the literature ($+$'s).  Points designated by diamonds ($\diamond$'s) represent predictions of work functions for which no values have appeared in prior literature. See part (c) of Table~\ref{tab:Results} for values plotted in this figure. The two regression lines in the graph are fit only to points designated by the $+$'s and derived from literature values of $W$.  Equations for the regression lines are given in the text of Section~\ref{sec:SecondApproxn}, along with a discussion of the two sets of linearly scaling points seen in the graph.}
\label{fig:CurvatureCorrectionBuckledResults}
\end{figure}
%

Further, the validity of this choice for $\delta$ and, therefore, of Eq.~(\ref{eq:basic_W_eqn}) is demonstrated here by the qualitative and quantitative accuracy of the results it so readily produces for $W$, using simply obtained estimates of $r_{WS}$.
This is seen for both planar and buckled 2D materials, respectively, in parts (a) and (b) of Table~\ref{tab:Results}, as well as in the graphs within Figs.~\ref{fig:PlanarResults} and \ref{fig:BuckledResults}.  Therein, one observes that the values Eq.~(\ref{eq:basic_W_eqn}) predicts ($\circ$'s in the graphs) are almost all within about 5\% or less of the literature values ($+$'s in the graphs).  The latter come either from experiment~\cite{rut2020graphene} or detailed, computationally intensive density functional theory (DFT) calculations~\cite{shahriar2022vacancy,thomas2020strain,zhang2022electronic,hasan2020strain,cui2018alkali,qin2012first,chen2016electronic,chen2016electronic2,garg2017band, jamwal2023tailoring,kripalani2018strain,sun2018exceptional,zhao2015new,shahriar2022adsorption}.

Additionally, in parts (a) and (b) of Table~\ref{tab:Differences} it is seen that almost all the individual values of $W$ predicted for both planar and buckled 2D materials, using the planar image-charge model and Eq.~(\ref{eq:basic_W_eqn}), are within the $\pm$0.3 eV error bound that has been attributed~\cite{DeWaele_etal2016} to DFT work function calculations.  The root-mean-square (RMS) differences from the literature for each of the two sets of values of $W$ calculated using Eq.~(\ref{eq:basic_W_eqn}) and presented, respectively, in parts (a) and (b) of Table~\ref{tab:Differences} are also well within that DFT error bound.

While Eq.~(\ref{eq:basic_W_eqn}) and the image-charge model for a planar conductor do work fairly well for $non$planar buckled 2D materials, as discussed immediately above, they provide only a first approximation for those materials, since they do not take explicit account of the out-of-plane structural features.  A second, even better approximation does do so by treating those features as though they introduce a local quasispherical curvature in a conductive material, to which a departing electron responds in a manner that is well-described via an image-charge model for a spherical conductive shell.  The equation for that second, more accurate, but still very algebraically simple electrostatics-based method is detailed below in Section~\ref{sec:SecondApproxn}.

%
\begin{table*}[t]
\caption{\small \small Parameter values for calculation of Wigner-Seitz radii $r_{WS}$ for 2D materials considered in this work.  The values are employed within the methods described in Section~\ref{sec:Methods} and yield values of $W$ given in Table~\ref{tab:Results}.  The number of mobile valence electrons for a hexagonal group of atoms is $N\!=\!6$ for all the 2D materials listed. Parameter $\lambda$ describes the extent of the overlap of the area $A$ covered by the distribution of those electrons with the region occupied by a neighboring hexagon, as depicted in Fig.~\ref{fig:2DMaterialsSchematic}(a).  Volume $V=\tau A$, where $\tau$ is the average thickness of the valence electron distribution and $\tau\!=\!r_B$ for single-atom materials, but $\tau\!=\!(1/2)r_B$ for two-atom materials. Values of lattice constants $a$ and bond lengths $\ell$ derive from DFT calculations described in the references cited in the last column of the table.  Materials designated by an asterisk have not had values for their work functions appear in the literature prior those listed in Table~\ref{tab:Results}.}
\label{tab:Parameters}
\begin{ruledtabular}
\begin{tabular}{cccccccccccc}
 &  &  &  &  &  & &  &  &Effective    &  \\
 &  &  &  & &  &  &  &   &Volume per  &Wigner-  &  \\
  & Lattice  &Bond & &Accessible   &Source  & Bohr  &  &       & Valence  &Seitz  &  \\
2D  & Constant,  &Length, &   &Area,   &of $r_{B}$  &Radius\footnotemark[1]&  Thickness,  & Volume  & Electron,  &Radius,  &  \\
Material  &$a$  &$\ell$ &$\lambda$ &$A$  &Value  &$r_{B}$  &$\tau$  &$V$  &${V
/N}$  &$r_{WS}$  &Ref.  \\
  &(Bohr)  &(Bohr) & &(Bohr$^2$)  &  &(Bohr)  &(Bohr)  & (Bohr$^3$) &(Bohr$^3$)  &(Bohr)  &  \\
\hline
\vspace{-2.5mm} \\
\multicolumn{12}{c}{\underline{(a) For Planar Structures}} \\
\vspace{-3mm} \\
Graphene  &4.65  &2.68  &1  &74.86  &$r_{B}$ of C  &1.276  &1.276  &95.52  &15.92  &1.561  &\cite{zhen2018structure,yang2018structure}  \\
2D BAs  &6.404  &3.698  &0.5  &79.92  &$r_{B}$ of As  &1.961  &0.981  &78.36  &13.06  &1.461  &\cite{shahriar2022vacancy}  \\
2D BN  &4.741  &2.737\footnotemark[2]  &0.5  &43.80  &$r_{B}$ of B  &1.653  &0.827  &36.20  &6.034  &1.129  &\cite{thomas2020strain}  \\
2D SiC  &5.86  &3.38\footnotemark[2]  &0.5  &66.87  &$r_{B}$ of Si  &2.092  &1.046  &69.95  &11.66  &1.407  &\cite{zhang2022electronic}  \\
2D BP  &6.070  &3.505  &0.5  &71.83  &$r_{B}$ of P  &1.754  &0.877  &63.00  &10.50  &1.358  &\cite{hasan2020strain}  \\
2D GaN  &6.149  &3.550\footnotemark[2]  &0.5  &73.68  &$r_{B}$ of Ga  &2.687  &1.344  &98.99  &16.50  &1.579  &\cite{cui2018alkali}  \\
2D TiN*  &6.085  &3.511  &0.5  &72.06  &$r_{B}$ of Ti  &2.975  &1.488  &107.20  &17.87  &1.622  &\cite{kokabi2023two}  \\
2D BSb*  &7.07  &4.08  &0.5  &97.33  &$r_{B}$ of Sb  &2.305  &1.153  &112.19  &18.70  &1.647  &\cite{elomrani2023evaluating}  \\
\vspace{-2mm} \\
\multicolumn{12}{c}{\underline{(b) For Planar Model of Buckled Structures Using Approximation 1, Eq. (\ref{eq:basic_W_eqn})}}  \\
\vspace{-3mm} \\
Silicene  &7.304  &4.303  &-  &31.43  &$r_{B}$ of Si  &2.092  &2.092  &65.74  &10.96  &1.378  &\cite{peng2013mechanical}  \\
Germanene  &7.79  &4.65  &-  &36.19  &$r_{B}$ of Ge  &2.245  &2.245  &81.25  &13.54  &1.479  &\cite{de2019first}  \\
Stanene  &8.857  &5.355  &-  &47.43  &$r_{B}$ of Sn  &2.587  &2.587  &122.71  &20.45  &1.696  &\cite{khan2017stanene}  \\
Plumbene  &9.32  &5.67  &-  &52.82  &$r_{B}$ of Pb  &2.793  &2.793  &147.52  &24.59  &1.804  &\cite{chaudhary2021future}  \\
Antimonene  &7.67  &5.44  &-  &41.76  &$r_{B}$ of Sb  &2.305  &2.305  &96.25  &16.04  &1.565  &\cite{mozvashi2020antimonene}  \\
Arsenene  &6.818  &4.739  &-  &32.31  &$r_{B}$ of As  &1.961  &1.961  &63.37  &10.56  &1.361  &\cite{shang2020permeability}  \\
BlueP\footnotemark[3]  &6.25  &4.33  &-  &27.07  &$r_{B}$ of P  &1.754  &1.754  &47.48  &7.913  &1.236  &\cite{shaikh2021ab}  \\
2D GaAs  &7.58  &4.52  &-  &34.22  &$r_{B}$ of Ga  &2.687  &1.344  &45.99  &7.664  &1.223  &\cite{bahuguna2016electric}  \\
2D SiGe*  &7.46  &4.44  &-  &33.15  &$r_{B}$ of Ge  &2.245  &1.123  &37.21  &6.202  &1.140  &\cite{nguyen2022chemical}  \\
2D GaP*  &7.28  &4.29  &-  &31.21  &$r_{B}$ of Ga  &2.687  &1.344  &41.94  &6.989  &1.186  &\cite{dange2023engineering}  \\
\vspace{-2mm} \\
\multicolumn{12}{c}{\underline{(c) For Nonplanar Model of Buckled Structures Using Approximation 2, Eq. (\ref{eq:Corrected_W_Eqn})}}  \\
\vspace{-3mm} \\
Silicene  &7.304  &4.303  &0.5  &108.23  &$r_{B}$ of Si  &2.092  &2.092  &226.41  &37.74  &2.081  &\cite{peng2013mechanical}  \\
Germanene  &7.79  &4.65  &0.5  &126.33  &$r_{B}$ of Ge  &2.245  &2.245  &283.61  &47.27  &2.243  &\cite{de2019first}  \\
Stanene  &8.857  &5.355  &0.5  &167.66  &$r_{B}$ of Sn  &2.587  &2.587  &433.74   &72.29  &2.584  &\cite{khan2017stanene}  \\
Plumbene  &9.32  &5.67  &0.5  &187.88  &$r_{B}$ of Pb  &2.793  &2.793   &524.74  &87.46  &2.754  &\cite{chaudhary2021future}  \\
Antimonene  &7.67  &5.44  &0.5  &173.15  &$r_{B}$ of Sb  &2.305  &2.305  &399.11  &66.52  &2.514  &\cite{mozvashi2020antimonene}  \\
Arsenene  &6.818  &4.739  &0.5  &131.31  &$r_{B}$ of As  &1.961  &1.961  &257.49  &42.92  &2.172  &\cite{shang2020permeability}  \\
BlueP\footnotemark[3]  &6.25  &4.33  &0.5  &109.47  &$r_{B}$ of P  &1.754  &1.754  &192.00  &32.00  &1.970  &\cite{shaikh2021ab}  \\
2D GaAs  &7.58  &4.52  &0.5  &119.24  &$r_{B}$ of Ga  &2.687  &1.344  &160.22  &26.70  &1.854  &\cite{bahuguna2016electric}  \\
2D SiGe*  &7.46  &4.44  &0.5  &115.28  &$r_{B}$ of Ge  &2.245  &1.123  &129.41  &21.57  &1.727  &\cite{nguyen2022chemical}  \\
2D GaP*  &7.28  &4.29  &0.5  &107.57  &$r_{B}$ of Ga  &2.687  &1.344  &144.53  &24.09  &1.792  &\cite{dange2023engineering}  \\  \\
\footnotetext[1]{Almost all Bohr radii $r_{B}$ given in the table are calculated as the reciprocals of values  of $\langle 1/r \rangle$ determined for the outermost valence atomic orbitals in the non-relativistic Hartree-Fock calculations of Bunge~\textit{et al.} \cite{bunge1993}.  An exception is the Bohr radius of Pb, which is calculated as the reciprocal of the value of ${\langle 1/r \rangle}$ determined in relativistic Dirac-Fock calculations by Visscher and Dyall \cite{visscher1997dirac}.}
\footnotetext[2]{Bond lengths $\ell$ of these footnoted planar 2D materials are calculated from their lattice constants $a$ via the geometric relationship ${\ell}$=$a$/$\!\sqrt3$, assuming that that the bonded atoms are in a regular hexagonal arrangement, as shown in Fig~\ref{fig:2DMaterialsSchematic}(a).  However, values of $\ell$ for other materials listed were obtained from the literature references cited in the last column of the table.}
\footnotetext[3]{BlueP is blue phosphorene, which is the buckled form of 2D phosphorus.}
\vspace{-8mm} 
\end{tabular}
\end{ruledtabular}
\end{table*}
%

For now, we simply observe that it produces the even more accurate predictions of $W$ shown in Fig.~\ref{fig:CurvatureCorrectionBuckledResults} and also in the bottommost portion, part (c), in each of Tables~\ref{tab:Results} and \ref{tab:Differences}.   In part (c) of Table~\ref{tab:Results} it can be seen that this second method for calculation of $W$ for buckled 2D materials produces values that are mostly within 2\% of those given by detailed DFT calculations of $W$ performed by others for the same materials~\cite{qin2012first,chen2016electronic,chen2016electronic2, sun2018exceptional,kripalani2018strain,shahriar2022adsorption,jamwal2023tailoring, zhao2015new}.  Also, in part (c) of Table~\ref{tab:Differences} the 0.11 eV RMS difference shown there for the second, nonplanar method is even smaller than the already small 0.14 eV RMS difference stated in part (b) for the accurate planar method, while in  Fig.~\ref{fig:CurvatureCorrectionBuckledResults}, most of the points representing the predictions from the second method appear to lie nearly on top of those based upon values from DFT calculations in the literature.

A key element in both of the above-described, electrostatics-based methods for predicting accurate values of $W$ are simple, physically-grounded procedures for accurately estimating values of $r_{WS}$.
Those procedures are detailed in the next section of this paper.

\section{\label{sec:Methods} Computational Procedures \\
and Results}

In general, the Wigner-Seitz radius is calculated~\cite{ashcroft1976solid}:   
\begin{equation}
\label{eq:WignerSeitzFormula}
r_{WS} = \bigg(\frac{3}{4 \pi}\frac{V}{N}\bigg)^{1/3} \,,
\end{equation}
where volume $V$ is the average volume accessible to an assembly of $N$ electrons.  The average area of the surface of a 2D material over which that volume extends is given here as $A$ and the distribution of the electrons is taken to have an average thickness $\tau$ over that area, so that
\begin{equation}
\label{eq:Volume}
V = \tau A \,.
\end{equation}

Recognizing that the 2D materials considered here consist of an enormous number of identical hexagonal structures, it is possible to determine these average properties for an entire material by determining them for just one hexagonal unit of six atoms in the surface of a 2D material, having lattice constant $a$ and bond length $\ell$.  Thus, here we take the $N$ electrons of interest to be the six mobile valence $\pi$-electrons associated with those six atoms---i.e., $N\!=\!6$.

\vspace{-0.4cm}
\subsection{\label{sec:Approximations} Two Simple Approximations}

In proceeding to estimate $A$, $\tau$, $V$, and thereby $r_{WS}$ for those six mobile valence electrons, one is estimating properties of the electron distribution over the 2D material.  Even standard complex methods, such as DFT, incorporate numerous approximations to estimate such electron distributions and their properties~\cite{burke2012perspective,hammes2017conundrum,li2020recent}.  Here, we make just two simple sets of approximations in order to determine $W$ with an accuracy comparable to that of DFT.   
We make these approximations based upon physical principles described and illustrated in much prior literature~\cite{wang2020planar,raty_etal2019,pople1967molecular,wahl1966molecular,wahl1970,bader1985atoms,hernandez2000properties,bunge1993,visscher1997dirac}

The first of these sets of approximations helps describe the extent of lateral electron delocalization of the valence electrons associated with a hexagonal group of atoms. For planar materials, this involves the introduction of a parameter $\lambda$ to estimate the fraction of a bond length $\ell$ along which the distribution of electrons from one hexagonal unit spreads into neighboring hexagonal units, as depicted in Fig.~\ref{fig:2DMaterialsSchematic}(a).  To keep our model simple, we assign $\lambda$ only two values that correspond to two key cases of interest.

For almost all the planar 2D materials discussed in this work we take $\lambda=1/2$.  This approximates the electron distribution from an atom in the hexagon of interest as extending about half way to the nucleus of an atom that is bonded to it, but which is outside the hexagon.  This is a common interpretation of electron sharing in covalent chemical bonds between atoms~\cite{raty_etal2019,fulton1993sharing}.  Solely for graphene, however, we take $\lambda=1$, which asserts that the valence electrons in one carbon atom are delocalized all the way to the position of the nucleus of a carbon atom bonded to it.  This larger value of $\lambda$ accounts for the much greater level of electron sharing and delocalization known to exist~\cite{raty_etal2019} within that all-carbon material, corresponding to its zero bandgap.

For buckled materials, though, delocalization and the planar area $A$ associated with the in-plane distribution of the valence electrons (i.e., those that sustain the image charge) can be thought of as being constrained and reduced by the portions of the hexagon that are bent out of the plane.  This is depicted in Fig.~\ref{fig:2DMaterialsSchematic}(b) and discussed in Section~\ref{sec:FirstApproxn}. 

The second set of approximations is applied to estimate $\tau$, the average thickness of the the electron distribution above the plane of the material.  To approximate this, we rely upon prior detailed studies of homonuclear and heteronuclear bonding~\cite{pople1967molecular,wahl1966molecular,wahl1970}.  These indicate that electrons are distributed relatively uniformly in a homonuclear bond and its mean radius is approximately the Bohr radius $r_B$ of each of the two identical bonded atoms.  This suggests that for all the 2D materials containing only one element or type of atom (i.e., ``single-atom materials") we take $\tau\!\approx\!r_B$, where $r_B$ is the reciprocal of $\langle 1/r \rangle$ for the valence $p$-shell of the atom, as given by \textit{ab initio} Hartree-Fock calculations~\cite{bunge1993,visscher1997dirac}.

These same bonding studies indicate that the electron distribution for a heteronuclear bond is asymmetrical and dominated by that for the larger atom, as are the dimensions of the bond.  This suggests that in the case of a 2D material containing two types of atoms (i.e., a ``two-atom material") the average thickness of the electron distribution should be estimated in terms of $r_B$ for the larger of the two atoms.  However, since only half the atoms in the material are of this larger species, we approximate the average thickness for the two-atom materials as $\tau\!\approx\!(1/2)r_B$.

Despite the anisotropy in the valence electron density distribution implicit in this model of a neutral two-atom 2D material, other investigators~\cite{thygesen2017calculating} have shown that, upon ionization from an anisotropic region of 2D material, a departing electron still will see an image charge delocalized throughout the surface of the material.  Thus, an image-charge model still should apply.  Here, the anisotropy just reduces $\tau$ and reduces $r_{WS}$, the distance above the neutral material from which an ionized electron is calculated to depart.   

The two simple, physically-motivated sets of approximations described above for estimating the mean dimensions of the valence electron distribution over a 2D material appear to be validated by the accuracy of the results they produce for $W$, as discussed in Section~\ref{sec:Intro}, when compared with the values produced by the DFT methods that are widely applied for that purpose.  In addition, we have performed sensitivity analyses to further validate our approximations and the simple sets of choices used here for values of $\lambda$ and $\tau$.

In these analyses, we found that varying the values of those parameters by $\pm 10\%$ altered the predicted values of $W$ by no more than about $\pm$0.01 Hartree---i.e., by no more than $\pm$0.27 eV, which is less than the estimated $\pm$0.3 eV error~\cite{DeWaele_etal2016} for a DFT calculation of $W$.  However, varying the parameters did diminish slightly the overall accuracy of the set of values calculated for $W$, relative to those from DFT---i.e., the RMS difference increased, but only by $\pm$0.002 Hartree or less.

Most important, though, when varying the values of $\lambda$ and $\tau$ and thereby altering the values of $r_{WS}$ corresponding to each of the 2D materials, the linear scaling of $W$ with $1/r_{WS}$ was maintained for their DFT values of $W$, along with those calculated here via Eq.~(\ref{eq:basic_W_eqn}).  That is, the central thesis of this work is not dependent upon the specific simple choices we have made to approximate the parameter values. 

Now, using the approximations introduced above, in order to calculate $r_{WS}$ via Eqs.~(\ref{eq:WignerSeitzFormula}) and (\ref{eq:Volume}), all that remains is to describe how to evaluate $A$ for the six valence $\pi$-electrons associated with a six-atom hexagon in the surface of a 2D material.  In the following two subsections we provide simple methods for doing so, using a planar model for both planar and buckled 2D materials.  In a third subsection, we take explicit account of the nonplanar nature of buckled 2D materials to derive a formula for $r_{WS}$ that helps determine even more accurate values of $W$ for those buckled materials than are given by the planar model and Eq.~(\ref{eq:basic_W_eqn}).

\vspace{-0.4cm}
\subsection{\label{sec:PlanarFormulas} Evaluating \textbf{\textit{\lowercase{r}\textsubscript{WS}}} to Determine \textbf{\textit {W} for Planar 2D Materials}}

For planar 2D materials, as mentioned above, we take the region of the planar surface for which we wish to determine the area $A$ to be the shaded hexagonal area shown in Fig.~\ref{fig:2DMaterialsSchematic}(a). We utilize for that purpose the bond length $\ell$ for the hexagonal arrangements of six atoms, as given in Table~\ref{tab:Parameters}.  
As is also seen in part (a) of the figure, the area of interest $A$ encompasses a single, primary hexagonally bonded structure in the surface of the material, 
plus the portions of adjoining hexagons into which $\pi$-electron density from the primary structure spreads.
Using the empirical parameter $\lambda$, introduced above to describe this overlap, the full hexagonal area available to the $\pi$-electrons (bounded by the dashed lines in Fig.~\ref{fig:2DMaterialsSchematic}(a)) will have a side of length $(1 + \lambda)\ell$, with the extended area 
\begin{equation}
\label{eq:PlanarArea}
A = \Big(\frac{3\sqrt3}{2}\Big)[(1 + \lambda)\ell]^2 \,.
\end{equation}
Then, by virtue of Eqs.~(\ref{eq:WignerSeitzFormula}) and (\ref{eq:Volume}), for the N=6 electrons of interest in a planar material:
\begin{equation}
\label{eq:Planar_rWS}
r_{WS} = \bigg\{\tau\Big( \frac{3\sqrt3}{16\pi} \Big) [(1 + \lambda)\ell]^2\bigg\}^{1/3}\,,
\end{equation}
with $\tau$ and $\lambda$ assigned values in the manner described above in Section~\ref{sec:Approximations}.

Using this expansion of $r_{WS}$, along with Eq.~(\ref{eq:basic_W_eqn}) and parameter values from part (a) of Table~\ref{tab:Parameters}, we obtain the values of $W$ listed in column 7 within part (a) of Table \ref{tab:Results} and graphed in Fig.~\ref{fig:PlanarResults}.  In that graph, as well as in Tables~\ref{tab:Results} and \ref{tab:Differences}, we once again note the accuracy of these $W$ values, as was also mentioned above in the Introduction.
In part (a) of Table~\ref{tab:Results}, values of $W$ predicted for planar 2D materials using the methods of this section are all seen to be within 5\% of literature values.  Further, from part (a) of Table~\ref{tab:Differences}, all the values predicted here for planar materials are within the $\pm$0.3 eV error bound that has been attributed~\cite{DeWaele_etal2016} to DFT work function calculations. 

\subsection{\label{sec:FirstApproxn} Planar First Approximation for $W$ \\in the Case of Buckled 2D Materials}

For $non$planar, buckled 2D materials, we may also apply a planar model and Eq.~(\ref{eq:basic_W_eqn}) to calculate $W$.  However, the equations for calculating $r_{WS}$ for use in Eq.~(\ref{eq:basic_W_eqn}) must be different from those applied for planar 2D materials in the previous section.
That is because, as is discussed in Section~\ref{sec:Approximations}, for the buckled materials we take the in-plane distribution of the six mobile valence electrons to be constrained to just the gray-shaded rectangular region of a hexagon that is depicted in Fig. \ref{fig:2DMaterialsSchematic}(b).  It has the area
\begin{equation}
\label{eq:BuckledArea}
A = a\ell  \, ,
\end{equation}
where $a$ is the lattice constant and $\ell$ is the bond length, as before.  Then, taking $N\!=\!6$, we have
\begin{equation}
\label{eq:Buckled_rWS_1}
r_{WS} = \bigg[\Big( \frac{\tau}{8\pi} \Big) a\ell\bigg]^{1/3}\,,
\end{equation}
from Eqs.~(\ref{eq:WignerSeitzFormula}) and (\ref{eq:Volume}), where the mean thickness $\tau$ is assigned values for single-atom and two-atom 2D materials as described in Section~\ref{sec:Approximations}.

Using Eq.~(\ref{eq:Buckled_rWS_1}) and Eq.~(\ref{eq:basic_W_eqn}), while taking values of $a$ and $\ell$ from part (b) of Table~\ref{tab:Parameters}, yields the values of $r_{WS}$, $1/r_{WS}$, and $W$ tabulated in part (b) of Table~\ref{tab:Results}.  The latter two sets of values are then plotted within the graph in Fig.~\ref{fig:BuckledResults}.
As seen in that table, those values for $W$ are almost all within 5\% or so of values from DFT published in the literature by other investigators.  In addition, from Table~\ref{tab:Differences}, all but one predicted value of $W$, that for GaAs, is within the $\pm$0.3 eV error range~\cite{DeWaele_etal2016} of those DFT-calculated literature values.

Despite this numerical accuracy, except for the $W$ values for silicene and germanene, values predicted for most of the single-atom buckled materials by Eq.~(\ref{eq:basic_W_eqn}) are seen in Fig.~\ref{fig:BuckledResults} to be systematically lower by a small amount than those from the DFT-calculated values in the literature. At the same time, the prediction for two-atom buckled 2D GaAs is much higher.
That is, while literature-derived points for most of the single-atom buckled materials are slightly above the dashed scaling line in the figure, the literature point for two-atom 2D GaAs is well below the line. (See Section~\ref{sec:Analysis} for a more in-depth discussion of this GaAs result.)

Nonetheless, in Fig.~\ref{fig:BuckledResults} a strong linear scaling trend versus $1/r_{WS}$ is still manifest in most of the points plotted for the literature values of $W$.  Also, using the nonplanar approximation presented in the next section, we can improve our predictions for $W$ relative to values of $W$ taken from the literature for  buckled materials.

\subsection{\label{sec:SecondApproxn} Nonplanar Second Approximation for $W$ \\in the Case of Buckled 2D Materials}

As accurate as are the values of $W$ predicted using the planar approximation of the preceding Section~\ref{sec:FirstApproxn}, by taking explicit account of the out-of-plane features of buckled 2D materials it is possible to obtain even more accurate predictions, as was suggested in the Introduction.  To do so, we use the methods outlined in Section~\ref{sec:PlanarFormulas} for determining $A$, $V$, and $r_{WS}$ in the case of a planar 2D material, but apply them to a buckled material.

\begin{figure}[t]
\begin{center}
\includegraphics[width=0.45\textwidth]{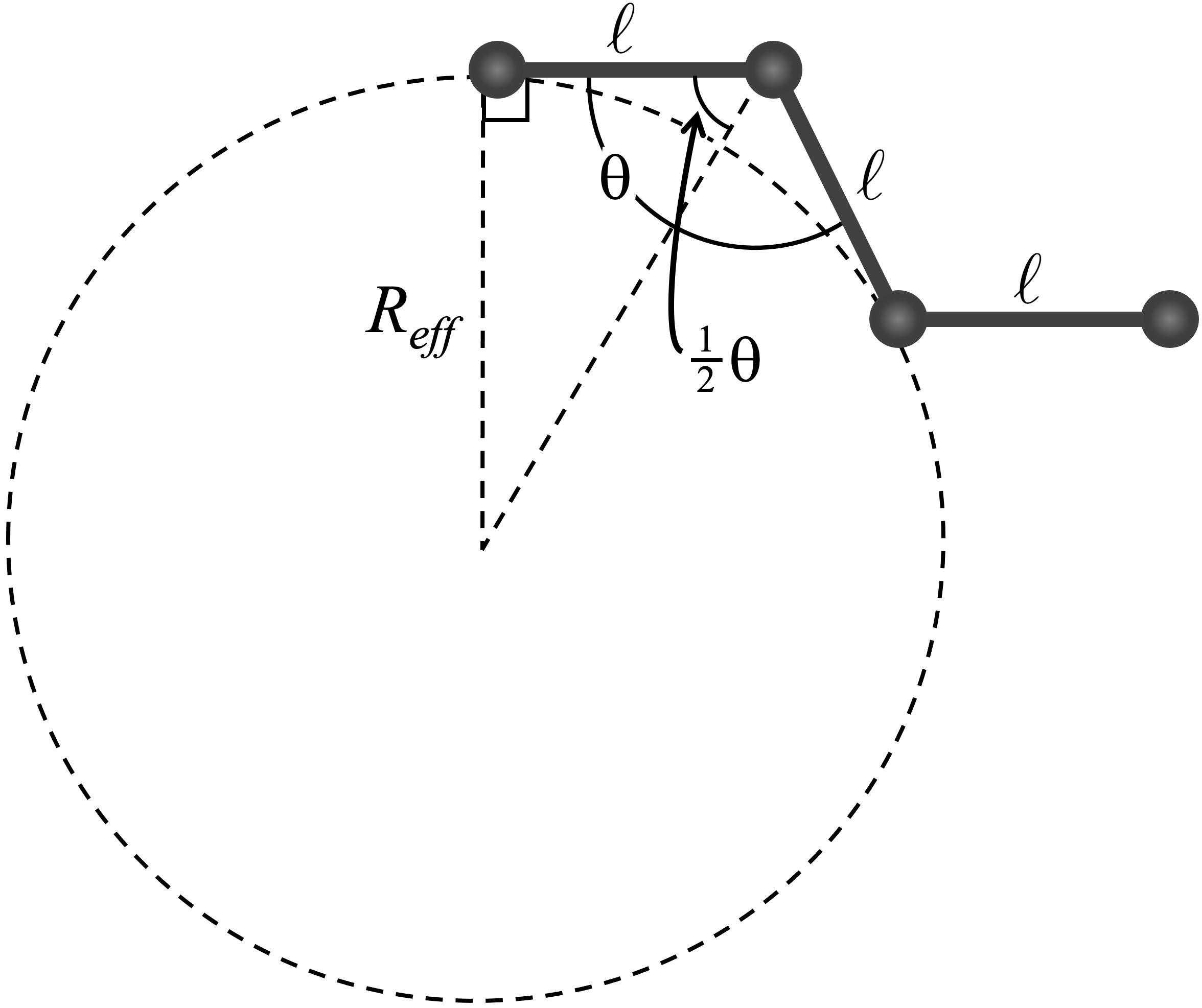}
\end{center}
\caption{\small \small Diagram illustrating how the bond length $\ell$ and bond angle $\theta$ characterize the ``bend" in the surface of a buckled 2D material and also how they figure into the determination of an effective radius $R_{eff}$ via Eq.~(\ref{eq:Effective_Radius}), which is used to calculate the material's work function via Eq.~(\ref{eq:Corrected_W_Eqn}).}
\label{fig:CurvatureCorrectionDiagram}
\end{figure}

\begin{figure}[t]
\begin{center}
\includegraphics[width=0.40\textwidth]{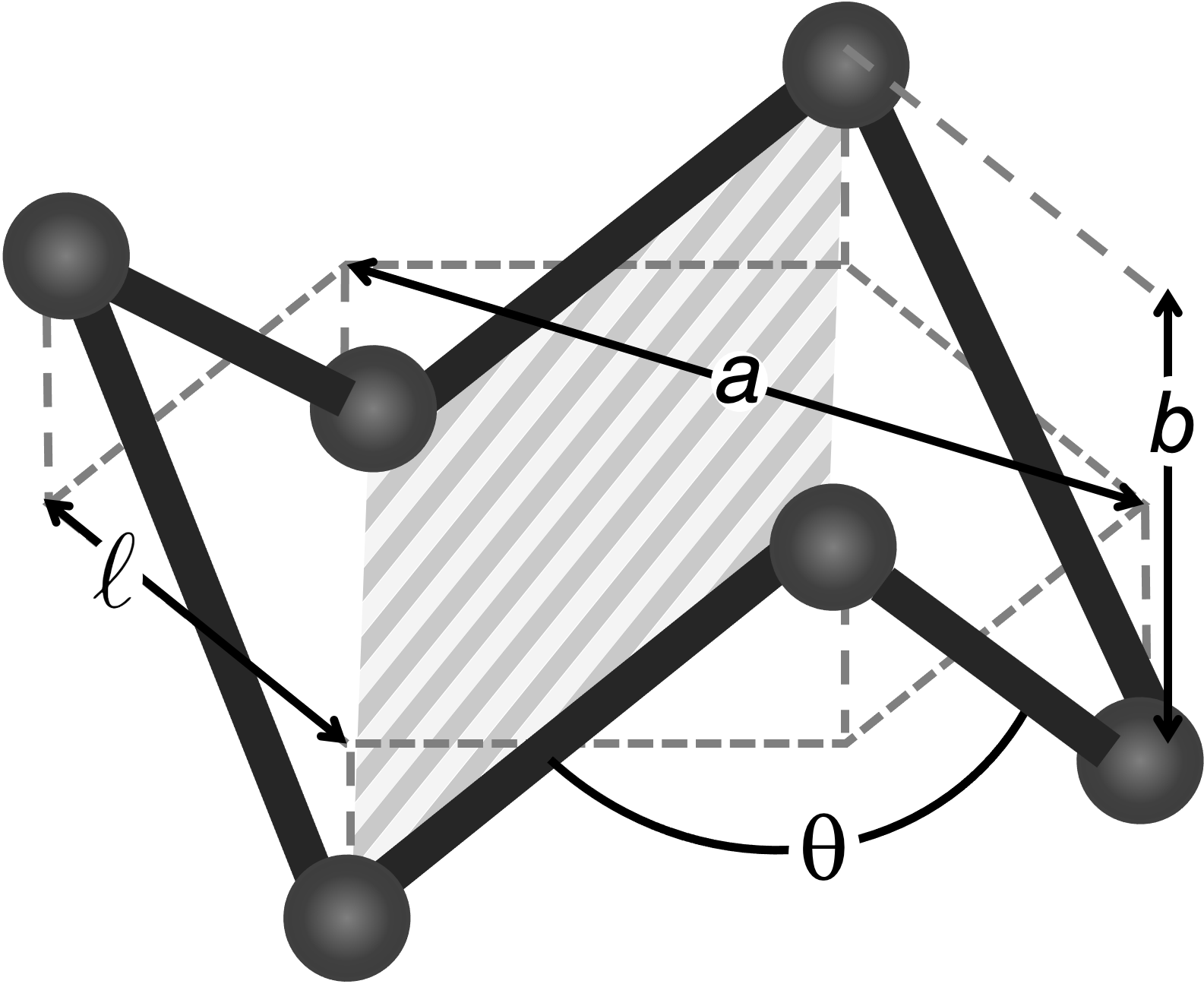}
\end{center}
\caption{\small \small Elevated side view of a 6-atom portion of a buckled hexagonal lattice in a buckled 2D material.  Dashed hexagon in the background is a projection of the buckled arrangement of atoms on a plane, illustrating the role of the lattice constant $a$, bond length $\ell$, and buckling height $b$ in characterizing the buckled structure.  The elevated gray-hashed rectangular region embodies the average area $A$, with dimensions $a$ by $\ell$, that is employed in the planar first approximation for calculating the work function of a buckled 2D material, as described in Section~\ref{sec:FirstApproxn}.  The figure illustrates, as well, the role of bond angle $\theta$ in determining the extent of the out-of-plane ``bend" in the material's surface where it buckles, of importance in the method of Section~\ref{sec:SecondApproxn}.  Figure is adapted from Ref.~\cite{grazianetti2016two}.}
\label{fig:Buckled_Structure}
\end{figure}

Then, to use the resulting value of $r_{WS}$ to determine $W$ for the buckled material, we employ a formula in which the first term is identical to the expression on the right-hand side of Eq.~(\ref{eq:basic_W_eqn}).  We add to it, though, an out-of-plane correction term, by treating the ``bend" in the material where it buckles as a section of a sphere.  For that purpose, we apply a spherical image-charge model~\cite{Landau_and_Lifschitz1984,jackson_electrdynamics2021} for the buckled surface that leads to an equation of the same form as one that has been applied to calculate work functions for small, spherical metal particles~\cite{Wood1981}:
\begin{equation}
\label{eq:Corrected_W_Eqn}
W = \frac{1}{4 r_{WS}} + \frac{3}{8} \frac{1}{R_{eff}} \,.
\end{equation}
Above, $R_{eff}$ is the effective radius of a sphere that can be inscribed in the the bend of a buckled region of a material's surface, as illustrated in Fig.~\ref{fig:CurvatureCorrectionDiagram}.  From that figure, trigonometry dictates that 
\begin{equation}
\label{eq:Effective_Radius}
R_{eff} = \ell\tan\Big(\frac{\theta}{2}\Big) \,,
\end{equation}
where $\theta$ is the angle between two neighboring bonds, as depicted in part (b) of Fig.~\ref{fig:2DMaterialsSchematic}.  Also, within Fig.~\ref{fig:Buckled_Structure} it is illustrated that this bond angle is equal to or a close approximation to the angle of the out-of-plane bend in the surface.  These angles are given in Table~\ref{tab:EffectiveRadius} for all the buckled 2D materials considered here, along with the corresponding values of $R_{eff}$ calculated via Eq.~(\ref{eq:Effective_Radius}).

Values of $W$ predicted by applying Eqs.~(\ref{eq:Corrected_W_Eqn}) and (\ref{eq:Effective_Radius}) are listed within part (c) of Table~\ref{tab:Results} and also plotted versus $1/r_{WS}$ in the graph within Fig.~\ref{fig:CurvatureCorrectionBuckledResults}.  In that part of Table~\ref{tab:Results} it is shown, as well, that most of the predicted values differ by only about 1\% or less from the values of $W$ published in the literature.  In part (c) of Table~\ref{tab:Differences} it is seen that all but one of those predicted values of $W$ differ by only a very small amount from the literature values and, taken together, they all have a small RMS difference, too.  As a result, most of the points in Fig.~\ref{fig:CurvatureCorrectionBuckledResults} representing values of $W$ predicted using the method of this section ($\circ$'s) lie almost on top of points representing the literature values ($+$'s).  (For a discussion of the less accurate points predicted for stanene and plumbene, see Section~\ref{sec:Analysis}.)

%
\begin{table}[t]
\caption{\small \small Parameters values used for the calculation of Effective Radius $R_{eff}$ via Eq.~(\ref{eq:Effective_Radius}).  Materials designated by an asterisk (*) have not had values for their work functions appear in the literature prior to the values listed for them in Table~\ref{tab:Results}.}
\label{tab:EffectiveRadius}
\begin{ruledtabular}
\begin{tabular}{ccccc}
  Buckled &Bond  &Bond  &Effective  &References  \\
  2D &Length,  &Angle,  &Radius,  &for  \\
  Material  &$\ell$  &$\theta$  &$R_{eff}$  &Dimensions  \\
  &(Bohr)  &(Deg.)  &(Bohr)  &  \\
\hline
\vspace{-3mm} \\
Silicene  &4.303  &116.1  &6.899  &\cite{peng2013mechanical}  \\
Germanene  &4.65  &111.81  &6.87  &\cite{de2019first}  \\
Stanene  &5.355  &111.4  &7.851  &\cite{khan2017stanene,cahangirov2016introduction}  \\
Plumbene  &5.67  &108.34  &7.85  &\cite{chaudhary2021future}  \\
Antimonene  &5.44  &89.6  &5.40  &\cite{mozvashi2020antimonene}  \\
Arsenene  &4.74  &91.97  &4.91 &\cite{shang2020permeability}  \\
BlueP  &4.33  &92.4  &4.51  &\cite{shaikh2021ab}  \\
2D GaAs  &4.52  &113.6  &6.90  &\cite{bahuguna2016electric}  \\
2D SiGe*  &4.44  &114.01  &6.84  &\cite{nguyen2022chemical}  \\
2D GaP*  &4.29  &115.575  &6.81  &\cite{dange2023engineering}  \\
\vspace{-4mm} \\
\end{tabular}
\end{ruledtabular}
\end{table}
%

Observe, as well, within Fig.~\ref{fig:CurvatureCorrectionBuckledResults} that points in the graph segregate themselves into two groups, each of which scales linearly with $1/r_{WS}$ along a separate regression line with a large $R^2$ value.  The upper and lower regression lines have the respective equations:
\begin{subequations}
\label{eq:RegressionEqns}
\begin{eqnarray}
W &= 0.36(1/r_{WS}) + 0.03  \label{eq:RegressionEqnsa}   \\
W &= 0.25(1/r_{WS}) + 0.06	\label{eq:RegressionEqnsb} \, .
\end{eqnarray}
\end{subequations}
These two regression lines are fit only to points derived from literature values of $W$ and designated by $+$'s.

However, since the predicted points designated by $\circ$'s lie so close to the literature points, as noted above, the predicted points also cluster into two groups along the two regression lines, as do the two points represented by diamonds ($\diamond$'s) that correspond to first-time predictions, for which there are no literature values.

In addition, we see that along the upper, dotted regression line, all the points correspond to materials composed of atoms from Group V of the periodic table.  Points along the lower, dotted and dashed regression line all correspond to materials composed of atoms from Group IV of the periodic table or combinations of atoms from groups III and V of the table (i.e., III-V materials).

Except for a small intercept term, Eq.~(\ref{eq:RegressionEqnsb}) for this lower regression line is nearly the same as Eq.~(\ref{eq:basic_W_eqn}), which describes scaling for planar 2D materials.
This similar trend arises from Eq.~(\ref{eq:Corrected_W_Eqn}) for two reasons. The first is that $r_{WS}$ is calculated in the nonplanar approximation in the same manner as for planar materials.
The second reason is that the out-of-plane correction term in $1/R_{eff}$ produces nearly constant values in the neighborhood of 0.05 for all the buckled group IV and III-V 2D materials, as shown in column 6 of Table~\ref{tab:Results}. Therefore, the scaling in $W$ with $1/r_{WS}$ comes almost entirely from the first term in Eq.~(\ref{eq:Corrected_W_Eqn}), which is identical to the right-side of Eq.~(\ref{eq:basic_W_eqn}) that governs the scaling for the planar materials.

%
%

%
%

By contrast, Eq.~(\ref{eq:RegressionEqnsa}) for the upper regression line in Fig.~\ref{fig:CurvatureCorrectionBuckledResults}, for the Group V 2D materials, has a larger slope than would be expected just from the first term of Eq.~(\ref{eq:Corrected_W_Eqn}).  As seen in columns 4 and 6 of Table~\ref{tab:Results}, this is because the three values from the second term increase progressively by the same amount, 0.06 Hartree, from the next smaller one as the approximately equally spaced values of with $1/r_{WS}$ increase.  As a result, the term $(3/8)(1/R_{eff})$, when plotted as a function of $1/r_{WS}$, would produce a line with a slope of approximately 0.13, from the values given in columns 3 and 6 of Table~\ref{tab:Results} for antimonene and BlueP.  Thus the slopes from the linear behavior of the first and second terms of Eq.~(\ref{eq:Corrected_W_Eqn}) add for the Group V materials to produce the still larger slope of 0.36 seen for the upper regression line in Fig.~\ref{fig:CurvatureCorrectionBuckledResults}.

Notable, as well, in Fig.~\ref{fig:CurvatureCorrectionBuckledResults}, it appears as though buckled 2D materials in each scaling group that are composed of larger atoms with larger atomic numbers also have larger values of $r_{WS}$ and, thus, smaller values of $1/r_{WS}$.  For this reason those materials appear toward the left along each scaling line and have smaller work functions, while materials that are composed of smaller atoms with lower atomic numbers appear progressively farther to the right and have both larger values of $1/r_{WS}$ and larger work functions.  This pattern also prevails in the results from the planar treatment of the buckled materials, described in Section~\ref{sec:FirstApproxn} and depicted in Fig.~\ref{fig:BuckledResults}.

The pattern is less clear, though, in work function results for the planar materials, described in Section~\ref{sec:PlanarFormulas} and depicted in Fig.~\ref{fig:PlanarResults}.  For one thing, the point for graphene appears well to the left within the graph in that figure, even though graphene is composed of relatively small carbon atoms.  This can be attributed to the greater delocalization of the electrons in graphene, as characterized by its larger value for $\lambda$, which compensates for the atom's smaller size.

\section{\label{sec:Analysis}Discussion and Analysis}

Any discussion or analysis of the foregoing technical presentation must begin by commenting upon the striking simplicity and apparent accuracy of the computational methods it describes, as well as upon the remarkable scaling regularities they reveal.  These scalings are seen to apply to work functions measured or calculated by others, as well as those calculated here.

Despite the extensive prior research and numerous publications focused upon 2D materials, as described in literature cited throughout this paper, it does not appear that the important scaling relationships that characterize the materials' electron detachment behavior have been noticed previously.  The existence of these relationships, depicted in Figs.~\ref{fig:PlanarResults}, \ref{fig:BuckledResults}, and \ref{fig:CurvatureCorrectionBuckledResults}, is an insight that would seem to be uniquely afforded by the electrostatics-based methods and viewpoint advanced in this work.

From the nature and origin of the formulas introduced and employed in this work, as well as their accuracy, it would seem that electrostatics dominates the electron detachment behavior in 2D materials, while quantum effects among the electrons would seem to play a more indirect role.  From this point of view, the main role of quantum mechanics in determining the magnitude of the electron detachment energies is to establish the bonding patterns and dimensions for the framework of atoms within the materials.  These govern the Wigner-Seitz radius, which then does govern the value of $W$. 

Of course, there are a variety of other structures and motifs for 2D materials~\cite{miro2014atlas,wang2020planar}, besides those treated here, with only hexagonal structures and with compositions including just one or two-types of atoms.  Most of those others do feature $\pi$-bonds between the atoms, though, with a distribution of valence electrons above the plane of the materials, in which an image charge can form and be delocalized to interact with a departing electron.  Thus, because of the seemingly primary role of this electrostatic effect in determining the energetics of electron detachment, the authors believe that Eq.~(\ref{eq:basic_W_eqn}) and the image-charge-based approach adopted here can be applied with success to materials with those other structures, as well.

Despite the remarkable accuracy of results predicted here by that approach, there are a few materials where the results are not as accurate as for most of the others.  One such case is the BN planar 2D material.  
The literature value~\cite{thomas2020strain} of $W$ for 2D BN, 0.211 Hartree, is the one in part (a) of Table~\ref{tab:Results} that is furthest, percentagewise, from the value 0.221 Hartree predicted here via Eq.~(\ref{eq:basic_W_eqn}), which is 4.7\% greater. The corresponding BN point in Fig.~\ref{fig:PlanarResults} is also the one furthest from the dashed $1/4r_{WS}$ scaling line dictated by Eq.~(\ref{eq:basic_W_eqn}).  It falls well below that line.
We attribute this to our assigning in Section~\ref{sec:Approximations} the parameter value $\lambda\!=\!1/2$ to describe delocalization of the valence electrons in BN.  This value works very well for most of the other materials.  However, molecular BN is isoelectronic with two bonded carbon atoms.  One might expect, therefore, that valence electrons in 2D BN behave a bit more like those in graphene and are more delocalized than those in other 2D materials.  

We tested this as part of the parametric sensitivity analysis mentioned in Section~\ref{sec:Approximations}. In fact, the larger value $\lambda\!=\!0.6$, corresponding to slightly greater delocalization, was found to produce a prediction $W\!=\!0.212$ Hartree via Eq.~(\ref{eq:basic_W_eqn}). This is within 0.001 Hartree or 0.027 eV of the DFT-calculated literature value, $W\!=\!0.211$ Hartree, for planar 2D BN.

Similar comments might be thought to apply for a number of the buckled 2D materials, as well, based upon their treatment using the planar approximation of Section~\ref{sec:FirstApproxn}, from which $W$ values differ from those in the literature by percentages as large as or larger than that for planar BN.  This is seen in part (b) of Table~\ref{tab:Results} and depicted in Fig.~\ref{fig:BuckledResults}.
However, it appears that these large disparities are not attributable so much to the parameter values, but to the the planar model or approximation for a buckled material.  We conclude this because the nonplanar model of Section~\ref{sec:SecondApproxn} gives much more accurate results for those same materials, using the same values for the parameters.  This is seen in part (b) of Table~\ref{tab:Results} and depicted in Fig.~\ref{fig:BuckledResults}.

Especially notable in this regard are the calculations for buckled 2D GaAs.  The $W$ value predicted using the planar model differs from the literature value~\cite{shahriar2022adsorption} by 7.9\%.  Since 2D GaAs is the only two-atom buckled material for which we show a literature value in part (b) of Table~\ref{tab:Results} and a point in Fig.~\ref{fig:BuckledResults}, this large disparity might be attributed to the planar model for a buckled material not providing a particularly good representation for such a two-atom material.  This view is reinforced by the fact that, while the planar first approximation of Section~\ref{sec:FirstApproxn} does not accurately predict the work function for GaAs, the nonplanar second approximation of Section~\ref{sec:SecondApproxn} calculates a value of $W$ for GaAs that is the same, to within 0.001 Hartree, as the DFT-calculated literature value.

Exceptions to the otherwise very accurate work function predictions from that nonplanar approximation are values calculated here for buckled 2D Sn and Pb, stanene and plumbene. This is apparent in the graph within Fig.~\ref{fig:CurvatureCorrectionBuckledResults}, as well as in part (c) of Tables~\ref{tab:Results} and \ref{tab:Differences}.  In the graph, points representing the stanene and plumbene predictions lie much below points representing the literature values.  Much as in the case of planar BN, this can be attributed to the fact that the $\lambda\!\!=\!\!1/2$ approximation does not work as well for these materials.  Unlike the BN case, though, for stanene and plumbene that approximation appears to $overestimate$ delocalization.  Electron sharing and delocalization is known to be less in Sn and Pb-based materials than for materials based upon lighter atoms from Group IV~\cite{raty_etal2019}.  This was tested in a further sensitivity analysis for $W$ as a function of $\lambda$.  It was found that $\lambda\!=\!0.35$ produced predictions of $W$ that differed by much less, no more than 0.003 Hartree (0.08 eV), from the literature values for both stanene and plumbene.

As can be inferred from the discussions above and in Section~\ref{sec:Methods}, the methods and calculations described in this work quickly and simply estimate and reveal the specific ways in which the electron distributions and energetics of 2D materials are sensitive to differences or changes in lattice dimensions, bond angles, out-of-plane-buckling or curvature, and composition.  This does require that one carefully assess the sensitivity of calculations to parameter choices, as we have done here.  (See above in this section and in Section~\ref{sec:Approximations}.)  However, the sensitivity of these easy-to-perform calculations to such changes in the structure and conformation of 2D materials might also be useful for assessing their impact, in advance, or even how to change properties for specific applications.  That is because a number of present and envisioned applications, especially for electronics, employ 2D materials in situations (e.g., stacked, bent conformationally, or stretched over a substrate or gap)~\cite{huang20222d,katiyar20232d} where their conformations are strained and compositions altered in just the ways to which the methods described here are sensitive.

\section{\label{sec:Summary}Summary and Conclusions}

To summarize, this paper has presented simple scaling rules that govern and accurately predict the values of work functions for both planar and buckled 2D materials.  The associated methods and formulas have been applied to a variety of such materials containing just one type of atom or two types.  The calculations were readily and quickly performed in a computer spreadsheet.

In cases where work function results have been published for the same materials previously, the results of the quick and easy calculations in this paper correspond very closely with those from  prior work.  However, all those prior results were much more difficult to obtain.  They involved either time-consuming experiments or computationally intensive DFT calculations.  The simple, electrostatics-based methods described here produce values of $W$ within the range of accuracy of those from prior DFT calculations.  In addition, we have predicted work functions values for several 2D materials that have not had any published previously.

Further, it is seen in this work that electrostatic scaling principles unexpectedly connect work function values that had been viewed previously in the literature as being discrete and unrelated to each other.  It is possible that these previously unappreciated scaling relationships among the detachment energies for a number of different 2D materials may provide insight into patterns that might apply to other properties of 2D materials, as well.

In future work, we plan to study other varieties of 2D materials, such as those with more different atoms, plus a greater variety of planar and nonplanar structures.  We hope to be able to treat the energetics of those systems with the same simplicity as in the work here, and possibly discover similar scaling regularities. 

\begin{acknowledgments}

The authors gratefully acknowledge valuable discussions with David Goldhaber-Gordon of Stanford University and valuable comments upon the manuscript by him and by Carl Picconatto, Shamik Das, and Jim Klemic of the MITRE Corporation. This research was supported by the MITRE Corporation.

\end{acknowledgments}


 \bibliography{Work_Functions_of_2D_Materials2025v18.bib}


\end{document}